\documentclass[ twocolumn,showpacs,preprintnumbers,amsmath]{revtex4}
\usepackage{epsfig}
\usepackage{graphicx}

\begin{document}

\title{Equilibrium under flow }

\author{F.~Gulminelli$^{(1)}$ and Ph.Chomaz$^{(2)}$ }

\affiliation{
(1) LPC Caen (IN2P3-CNRS/ISMRA et Universit\'{e}), F-14050 Caen C\'{e}dex,
France \\
(2) GANIL (DSM-CEA/IN2P3-CNRS), B.P.5027, F-14021 Caen C\'{e}dex, France }

\begin{abstract}
{\small Using information theory we derive a thermodynamics for systems
evolving under a collective motion, i.e. under a time-odd constraint. 
An illustration within the Lattice gas
Model is given for two model cases: a collision between two complex
particles leading to a incomplete relaxation of the incoming momentum, and a
self-similar expansion. A semi-quantitative connection with the
determination of thermodynamical quantities in multifragmentation reactions
 is done showing that they are affected in a sizeable way only when the 
 flow dominates the global energetic. }
\end{abstract}


\maketitle


Heavy ion collisions represent a unique opportunity to probe
the interdisciplinary field of phase transitions in mesoscopic systems for
which a non standard thermodynamics giving rise to negative heat capacities
is predicted 
\cite{cneg_theo}.  An indication of 
negative heat capacities has been experimentally found 
\cite{cneg_exp} in 
multifragmentation reactions. These results can be considered as a first
step towards a quantitative determination of the nuclear phase diagram.
Negative heat capacities have also been observed in simulations of
self-gravitating systems\cite{grav} and in the melting and boiling 
of clusters \cite{haberland}.

An important conceptual problem linked with multifragmentation
experiments is that the outcomes of a nuclear collision are not confined in
an external container but dynamically deexcite in the vacuum. In the absence of
boundary conditions, collective flows 
can be present at the fragmentation time. Because
of this time odd component, one may doubt about the applicability
of equilibrium concepts. However, we will show that information theory allows
a thermodynamically consistent description of open finite systems in
evolution under a collective flow. We will apply this formalism to the
Lattice Gas Model \cite{LGM} for the particular cases of a memory of the
entrance channel (transparency) and
radial expansion. In both cases the properties of the system 
appear to be affected in a sizeable way when flow
dominates the global energetics.


In the Gibbs formulation of statistical mechanics an equilibrium is defined
as a collection of different microstates all corresponding to the same
macrostate (or statistical ensemble) defined through the average value of 
(a collection of) collective variables. A practical realization of a Gibbs
ensemble is given by the collection of different snapshots of an isolated
ergodic system evolving in time, provided that  
the observation time is much longer than a typical equilibration time
. This type of statistical ensemble is not very
useful for short-lived open systems as the transient excited states formed
in a nuclear collisions. However, ergodicity is not the unique way to
produce a Gibbs ensemble 
. Indeed, within information theory, an
equilibrium corresponds to any ensemble of states that maximizes the entropy
in a given space under the constraint of (a number of) observables known
in average\cite{balian}. Therefore, if the nuclear dynamics is sufficiently
sensitive to the initial conditions, the ensemble of outcomes of
similarly prepared nuclear collisions can be considered as a statistical
ensemble for which the important observables are controlled by the dynamics 
(and by the event sorting performed on specific observables).
As an example, in the standard freeze out hypothesis the configurations are
fixed when the nuclear interaction among prefragments becomes
negligible; then the ensemble of events is
characterized by several variables such as the energy
and the spatial extension 
$R^{2}$ which for open systems is fluctuating event by event. 
This can be accounted for by considering the ensemble average $%
<R^{2}>$ as a state variable and introducing a Lagrange multiplier $\lambda $
closely related to a pressure. In this information theory approach, the state
variables are determined by the dynamics and
can also be time odd quantities such as transparency or
radial flow. 


Let us first consider a symmetric head-on collision with a too short reaction time
to fully relax the incoming momentum. This situation seems to be verified at
relativistic energies \cite{rami}. 
It corresponds to the observation 
of an additional one body
state variable, the memory of the initial momenta $<\epsilon p_{z}>$, where $%
p_{z}$ is the momentum along the beam axis and $\epsilon =-1 (+1)$ 
for the particles initially belonging to the target (projectile). 
Let us assume that the total energy $E$ is also known only in average.
The maximization of  the
entropy leads to the partition sum of the incomplete momentum relaxation
ensemble (IMRE)
\begin{equation}
Z_{\beta ,\alpha }=\sum_{n}exp\left( -\beta E^{(n)}+\alpha
\sum_{i=1}^{A}\epsilon _{i}{p_{iz}^{(n)}}\right)   \label{zeta}
\end{equation}
where {the} indice $i$ stands for the $i$-th particle while $%
(n)$ counts the events. $\beta $ and $\alpha $ are Lagrange multipliers 
associated to the constraint of the
degree of incomplete stopping $<\epsilon p_{z}>$ and of 
the total center of mass energy $<E>$.  
The relative probability of an event $(n)
$ results
\begin{equation}
p^{(n)}\propto exp\left[ -\beta \left( \sum_{i=1}^{A}\frac{\left( \vec{p}%
_{i}^{(n)}+\epsilon _{i}\vec{p}_{0}\right) ^{2}}{2m}+\sum_{ij}^{A}U_{ij}%
\right) \right]  \label{prob1}
\end{equation}
where $U_{ij}$ is the two body
interaction and we have introduced
  $\vec{p}_{0}=m\alpha/\beta \vec{u_z} $%
. The average kinetic energy is 
$<E_{k}>/A=3/(2\beta)+ p_{0}^{2}/ (2m) $  
while the  equation of states related to $\alpha $ leads to 
$<\epsilon p_{z}>=Ap_{0}$.
In the limit $p_{0}=0,$  these expressions correspond  to the
usual  canonical ensemble while in the general case they can be 
interpreted as two thermalized sources with a
non zero relative velocity $2\vec{p}_{0}/m$ 
along the beam axis $\vec{u_z}$ 
. 
\begin{figure}[tbh]
\includegraphics[height=0.9\linewidth]{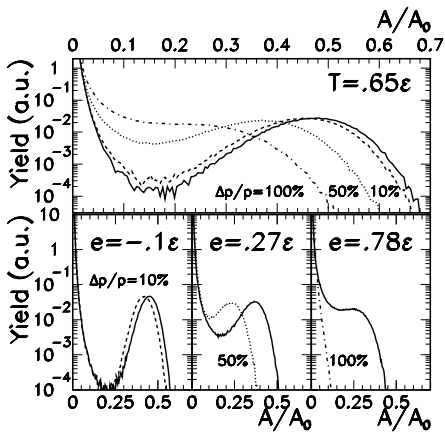}
\caption{\textit{Fragment size distributions in the IMR Lattice Gas model. 
Upper part: constant temperature plus an
increasing proportion of directed linear momentum.Lower part: microcanonical
calculations. Full lines: canonical and microcanonical calculations with an
isotropic momentum distribution. }}
\label{f1}
\end{figure}
In actual heavy ion experiments the centrality selection criteria imply a
sorting of data according to the total deposited energy \cite{cneg_exp} or
to variables which are strongly correlated to it. This means that a total
energy conservation has to be implemented to eq.(\ref
{prob1}). In this case the equations of state 
are not analytical but can still be numerically evaluated.

This theory can be extended to take into account other observables. 
 if for example the average momentum dispersion along the beam axis is known, 
an extra term $\delta p_{iz}^{(n)2}$ can  be added in eq.(\ref{zeta})
to control the collective flow fluctuation, while a term $\lambda R^{(n)2}$
can be used to impose an average freeze out volume.
For simplicity in the following calculations the longitudinal momentum 
dispersion has been kept
fixed, $(\beta/2m+\delta)^{-1}=0.04\epsilon $.

To understand the effect of transparency on the evaluation of
thermodynamical quantities we have used the Lattice Gas hamiltonian \cite
{LGM} where occupied sites on a three dimensional cubic lattice interact via
a constant closest neighbors coupling $\epsilon $. This simple but
numerically solvable model is isomorphous to the Ising model in the
grancanonical ensemble and constitutes therefore a paradigm of standard
equilibrium statistical mechanics with first and second order phase
transitions. Calculations in the IMRE eq.(%
\ref{zeta}) are performed for a system of $A=216$ particles at a subcritical
pressure $\lambda =3.3\cdot 10^{-4}$  and
clusters are defined within the standard Coniglio-Klein
prescription\cite{dasgupta}%
. Since in this classical approach the configurational and kinetic partition
sums are factorized, for a given $\beta$ the lattice
configurations will be independent of the degree of transparency $p_{0}$. 
However the active bond probability will
explicitly depend on $p_{0}$ meaning that cluster observables can be
affected by the incomplete relaxation. 
\begin{figure}[tbh]
\includegraphics*[height=0.55\linewidth, trim=100 250 100 50]{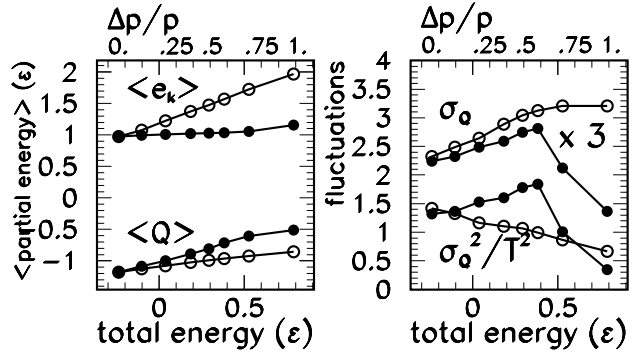}
\caption{\textit{Average partial energies (left) and variances (right) in
the microcanonical ensemble (full circles) and in the IMRE 
(open circles) at fixed energy. }}
\label{f2}
\end{figure}
Figure 1a shows the cluster size distributions for a temperature $\beta
^{-1}=0.65\epsilon $ which corresponds to the transition temperature in the
canonical ensemble, and different quadrupolar deformations in momentum space 
\begin{equation}
\frac{\Delta p}{p}=\frac{\sqrt{2}<|p_{z}|>-<|p_{\perp }|>}{2(\sqrt{2}%
<|p_{z}|>+<|p_{\perp }|>)}  \label{deltap}
\end{equation}
 A quantitative comparison with experimental data would require to fix
the Lagrange parameters $\lambda $, $\beta $, $p_{0}$ from each specific set
of data and to include quantum effects and symmetry as well as Coulomb terms
in the hamiltonian, however from this illustrative example we can clearly
see that partitions are 
affected by a collective longitudinal
component and a higher degree of fragmentation does not necessarily imply
higher temperatures but can also be consistent with an increased degree of
transparency of the collision. To quantify this statement, Figure 1b
compares the clusters size distributions of the IMRE
 at a fixed total energy with the standard microcanonical
ensemble (fixed energy, spherical momentum distribution) at the same energy.
It is clear that thermal agitation is much more effective than transparency
to break up the system: at the energy where the microcanonical ensemble
predicts a complete vaporization of the system, a residue persists
if the non relaxed momentum component is as large as the relaxed one. 
This is in qualitative
agreement with the trend observed in central collisions at relativistic
energies \cite{fopi}. On the other side up to 10\% transparency the
distributions are almost identical, meaning that when the velocity
difference between the quasi-projectile and the quasi-target is of this
order, the debate on equilibrium based on the number of emission sources 
\cite{indra} is an academic question. 
The important influence of the collective component on the cluster
distributions means that, if the incoming
momentum is not completely relaxed, the methods used to determine 
thermodynamical quantities can be strongly biased. Indeed if energy is
divided into a kinetic $e_{k}$ and an interaction $e_{i}$ component, the
average kinetic energy can be used as a microcanonical thermometer while the
normalized partial energy fluctuations are linked to the microcanonical heat
capacity \cite{noi}. The interaction energy can be
estimated as the Q-value of the detected fragments 
\cite{cneg_exp}. 
Figure 2 shows the deformation induced by transparency 
on the first and second order moments of partial energies that are used to
infer the temperature and heat capacity in the microcanonical sorting of
fragmentation data \cite{cneg_exp}. The Q-value can be computed in the
Lattice Gas model in a liquid drop approximation as 
\begin{equation}
Q^{(n)}=\sum_{i=1}^{M^{(n)}}a_{v}A_{i}+a_{s}A_{i}^{2/3}  \label{qvalue}
\end{equation}
where $M^{(n)}$ is the multiplicity of the n-th event and $a_{v}$, $a_{s}$
are the volume and surface energy coefficients. The average kinetic energy
and the average Q-value obtained with
eq.(\ref{qvalue}) are shown in the left part of figure 2 for the
microcanonical ensemble and for a system with the same total deposited
energy and an increasing degree of transparency $\Delta p/p$. The energy
range explored corresponds to the phase transition region. The increased
probability of heavy clusters with an increasing degree of transparency
shown in figure 1 induces an overestimation of the average kinetic energy,
hence of the estimated temperature. The peak in the partial energy
fluctuation which signals the phase transition \cite{noi} disappears in this
extreme scenario of an increase of energy going entirely into the relative
motion of the two sources, and the normalized fluctuations are systematically
lower than in the microcanonical ensemble where the incoming momentum is
fully relaxed. Since a negative heat capacity corresponds to abnormally high
partial energy fluctuations \cite{noi}, Figure 2 implies that an incomplete
relaxation of the incoming momentum can prevent the observation of 
negative heat capacity. 


%
%
\begin{figure}[tbh]
\includegraphics[height=.75\linewidth]{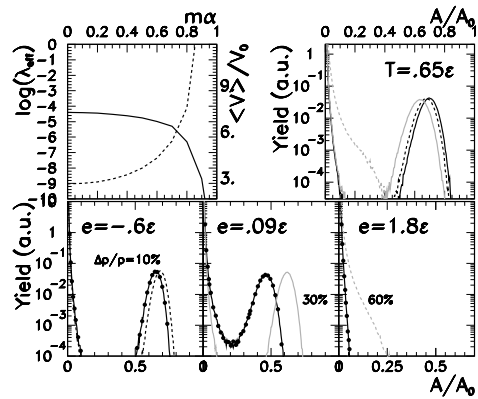}
\caption{\textit{Upper left: effective pressure (full line) and average
volume (dashed line) as a function of the collective radial velocity. Upper
right and lower part: fragment size distributions in the expanding Lattice
Gas model. Distributions without flow (full black lines) are compared with
distributions with 10\% (dashed black), 30\% (full grey) and 60\% (dashed
grey) contribution of radial flow at the same temperature (upper right) and
at the same total energy (low part). Symbols: calculations with flow at the
same thermal energy and average volume as the full black lines. }}
\label{f3}
\end{figure}
Another important form of collective motion in heavy ion collisions is
radial flow which starts to be observed in central collisions around 30
A.MeV incident energy\cite{indra} and becomes the dominant fraction of the
detected energy in the relativistic domain \cite{fopi}. 
We can describe this dynamical situation in the framework of information
theory as an equilibrium with non random directions for velocities which are
preferentially oriented in the radial direction. In the canonical
formulation this corresponds to two independent observations of the average
energy $<E>$ and a the average local radial momentum $%
<p_{r}(r)>$ . The probability of a microstate $(n)$ reads 
\begin{equation}
p^{(n)}\propto exp\left( -\beta E^{(n)}-\sum_{i=1}^{A}\gamma (r_{i})\vec{p}%
_{i}\cdot \vec{u}_{r_{i}}\right)   \label{prob2}
\end{equation}
where $r_{i}$ is the center of mass position of the i-th particle and $\beta
,\gamma \left( r\right) $ are Lagrange multipliers. Imposing in the local
equation of state $<p_{r}>=\partial logZ/\partial \gamma $ that the observed
velocity is self-similar $<p_{r}(r)>=m\alpha r$ we obtain $\gamma (r)=-\beta
\alpha r$ which gives for the argument of the exponential in the probability
(\ref{prob2}) 
\begin{equation}
-\beta \sum_{i=1}^{A}\frac{\left( \vec{p}_{i}-\vec{p}_{0}(r_{i})\right) ^{2}%
}{2m}+\beta \frac{\alpha ^{2}m}{2}\sum_{i=1}^{A}r_{i}^{2}-\beta E_{pot}
\label{prob3}
\end{equation}
with the local radial momentum $\vec{p}_{0}(r)=m\alpha r\vec{u}_{r}$. The
situation is equivalent to a standard Gibbs equilibrium in the local
expanding frame. This scenario is often invoked in the literature \cite{smm}
to justify the treatment of flow as a collective radial velocity
superimposed on thermal motion; however eq.(\ref{prob3})contains also an
additional term $\propto r^{2}$ which corresponds to an outgoing
pressure. This term cannot be avoided in the selfsimilar scenario; indeed
the total average energy under flow reads
$ <E>={3}/({2\beta })+<E_{pot}>+ m\alpha
^{2}/2<R^{2}> $ 
where we have used the fact that the scalar product between the flow
and the thermal component is in average
zero. The probability under flow eq.(\ref{prob3}) diverges at infinity
reflecting the trivial dynamical fact that asymptotically particles flow
away. This divergence should be cured by introducing an external confining
pressure which is not a mathematical artifact but has to be
interpreted as discussed above as 
a Lagrange multiplier imposing a finite freeze out volume. Eq.(\ref{prob3})
has then to be augmented by a term $-\lambda \sum_{i}r_{i}^{2}$ with $%
\lambda \geq m\alpha ^{2}/2T$ leading to a positive pressure
coefficient $%
\lambda _{eff}=\lambda -m\beta \alpha ^{2}/2$.  


The effective pressure $\lambda _{eff}$ as well as the associated average
volume (normalized to the ground state volume $V_{0}=A$) are shown in the
upper left part of figure 3 as a function of the collective radial velocity
for a given pressure $\lambda =1.23\cdot 10^{-2}$ and temperature $\beta
^{-1}=0.65\;\epsilon $. The Lagrange parameter $\lambda _{eff}$ being a
decreasing function of $\alpha $, the flow reduces the effective pressure
  so that the critical point is moved towards higher pressures in the
presence of flow \cite{jou}. However one can see that the effect is very
small up to $m\alpha \approx .6$ (which corresponds to $\approx 40\%$
contribution of flow to the kinetic energy). In this regime the
cluster size distributions displayed in the upper right part of figure 3 are
only slightly affected. On the other side if the collective flow overcomes a
threshold value $\Delta E_{k}/E_{k}\approx 50\%$ the average volume (dashed
line in fig.3a) shows an exponential increase and the outgoing flow pressure
leads to a complete fragmentation of the system (dashed grey line in fig.3b).
Again
an oriented motion is systematically less effective than a random one to break
up the system. This is shown in the lower part of figure 3 which compares
for a given $\lambda$ distributions with and without radial flow at the same
total deposited energy: for any value of radial flow equilibrium in the
standard microcanonical ensemble corresponds to more fragmented
configurations.

Concerning heavy ion collisions, it is important to stress that the pressure 
$\lambda $ as well as the other state variables
are consequences of
the dynamics. They cannot be accessed by a statistical treatment but have to
be extracted from simulations and/or directly infered from data itself \cite
{cneg_exp}. Different models assume that fragmentation occurs in an average
freeze out volume which may depend on the thermal energy but does not depend
on flow. This is true if the system fragments at the turning point of its
expansion ($\lambda _{eff}=0$)\cite{papp} or when the interaction between
fragment surfaces becomes negligible \cite{smm} or more generally
insufficient to modify the N-body correlations \cite{sator}. In this case
the presence of flow does not affect the configuration space and can only
modify the partitions because of the modified microscopic bonds among
particles taken into account by the Coniglio-Klein algorithm. However this
effect is negligible as already observed by Das Gupta et al.\cite{dasgupta}
and shown in the lower part of Figure 3. In this figure the symbols
represent the size distributions with collective flow at the same thermal
energy and average volume as the standard microcanonical results (full
lines). Even for the largest amount of flow considered the two distributions
are identical, meaning that in this hypothesis 
all thermodynamical analysis of
fragmentation data stay valid in the presence of even strong collective flows
\cite{dasgupta}.


In this paper we have shown that information theory \cite{balian} allows to
take into account generic collective motions in a rigorous statistical way
leading to the formulation of a consistent ''out of (static) equilibrium''
thermodynamics. We have particularized this general statement to two kinds
of collective motion which are particularly relevant in nuclear
multifragmentation, namely partial transparency and radial selfsimilar flow.
We have applied the formalism to the Lattice Gas model which 
has already shown some pertinence \cite{dasgupta} to the
multifragmentation phenomenology. We have shown that the disordered thermal
motion is always more efficient than the collective motion to break up the
system. A consequence of that is that normalized partial energy fluctuations
for a system under flow are always reduced respect to the standard
microcanonical expectation\cite{cneg_theo} and the negative heat capacity
signal experimentally observed\cite{cneg_exp} cannot be due to the possible
presence of a collective component neglected in the data analysis.
Transparency leaves the configurational energy unmodified, while radial
flow, because of the coupling between r and p space, is shown to be
equivalent to an external pressure leading to an increased value of the
critical point\cite{jou}. Concerning the extraction of thermodynamical
variables from fragmentation data, radial flow can be simply subtracted from
the detected energy \cite{dasgupta} while transparency can be neglected only 
if the relaxation of the incoming momentum exceeds about 90\%.

\end{document}